\documentclass[english]{article}
\usepackage[T1]{fontenc}
\usepackage[latin9]{inputenc}
\usepackage{amsmath}
\usepackage{amssymb}
\usepackage{graphicx}

\makeatletter
\newcommand{\lyxaddress}[1]{
	\par {\raggedright #1
	\vspace{1.4em}
	\noindent\par}
}

\usepackage{amsfonts}

\makeatother

\usepackage{babel}
\begin{document}

\title{Self-assembly of core-corona particles confined in a circular box}

\author{Erik R. Fonseca and Carlos I. Mendoza\textsuperscript{}\thanks{E-mail: cmendoza@materiales.unam.mx}\\
}
\maketitle

\lyxaddress{Instituto de Investigaciones en Materiales, Universidad Nacional
Autónoma de México, Apdo. Postal 70-360, 04510 CdMx, Mexico}
\begin{abstract}
Using Monte Carlo simulations, we study the assembly of colloidal
particles interacting via isotropic core-corona potentials in two
dimensions and confined in a circular box. We explore the structural
variety at low temperatures as function of the number of particles
($N$) and the size of the confining box and find a rich variety of
patterns that are not observed in unconfined flat space. For a small
number of particles $\left(N\leq6\right)$, we identify the zero-temperature
minimal energy configurations at a given box size and we construct
the phase diagram as function of temperature and box radius for the
specific case of $N=4$. When the number of particles is large ($N\geq100$),
we distinguish different regimes that appear in route towards close
packing configurations as the box size decreases. These regimes are
characterized by the increase in the number of branching points and
their coordination number. In contrast to the case of confined hard
disks, we obtain open structures with unexpected highly anisotropic
character in spite of the isotropy of the interactions and of the
confinement. Our findings show that confined core-corona particles
can be a suitable system to engineer particles with highly complex
internal structure that may serve as building blocks in hierarchical
assembly.
\end{abstract}

\section{Introduction}

The search of particles on the mesoscopic scales that self-organize
into potentially useful structures by virtue of their mutual interactions
is extremely important since they can be used as building blocks for
bottom-up nanofabrication processes \cite{rechtsman1,rechtsman2}
. This is a large and rapidly growing field of tremendous technological
potential and fundamental interest that has been fed by the continuing
progress in the manipulation of the interaction potentials between
nanoparticles \cite{min}-\cite{tang}.

Successful self-assembly of a structure depends not only on the interactions
between the components as other factors intervene, for example reversibility,
to allow the components to adjust their position within a structure,
the thermal motion usually required to provide movement to the components
and the environment where the components move.

In past decades, it has been recognized that one important factor
affecting the assembly of particles is confinement. A significant
amount of scientific attention has been devoted into the self-assembly
of colloidal systems in different types of confinement. This interest
is partly due to important problems in biosciences that involve objects
confined inside a cavity \cite{wan}. To mention just a few, DNA packaging
in viral capsids \cite{marenduzzo}, macromolecular crowding in the
cell \cite{ellis}, blood clotting \cite{cines} and pattern formation
in biological structures \cite{hayashi}. Additionally, systems of
physical relevance like the growth of colloidal crystals under confinement
\cite{velev}-\cite{manoharan3}, the transport of particles in channels
\cite{burada,reguera} and others have contributed to the interest
in studying the arrangement of particles under these conditions. As
a result, the study of confinement has been used not only as a way
to understand the behavior of the confined objects but as a general
way of tuning colloidal self-assembly \cite{sankaewtong}-\cite{Schmidt}.

Most monodisperse colloidal particles are spheres. Thus, one problem
in building new mesoscale ordered materials is controlling how spheres
pack. Confinement of particles within containers such as micro-patterned
holes or spherical droplets produce unexpected polyhedra that may
become building blocks for more complex materials \cite{manoharan1,manoharan2}.

In most cases, colloidal crystals found in confinement are close-packed.
This can be a desired characteristic in problems related to efficient
manufacturing, packaging, and transport \cite{teich}. However, one
of the major goals in colloidal self-assembly is to devise new colloidal
systems to self-assemble into low-density open crystalline structures
\cite{zhu,salgado-blanco} which are partially due to their promising
applications in photonics \cite{galisteo-lopez}, catalysis \cite{stein},
porous media \cite{su} and the special response to mechanical stress
\cite{souslov}-\cite{mao}. Assemblies in the form of two-dimensional
open networks are also of particular interest for possible applications
because well-defined pores can be used for the precise localization
and confinement of guest entities such as molecules or clusters, which
can add functionality to the supramolecular network \cite{madueno}.
Two-dimensional packings of hard spheres under circular confinement
results in optimal circle packings forming doublets, triangles, squares,
pentagons, and hexagons \cite{kravitz}. It has been recently demonstrated
by numerical simulations that soft particles described by the Daoud-Cotton
model for star polymers assemble open structures in two dimensions
\cite{zhu}. Here we use computer simulations to investigate the self-assembly
in two dimensions of particles consisting of a hard core surrounded
by a soft corona that are confined in a circular box and the possibility
to use confinement as a tool to control the final self-assembled structure,
in particular, to obtain open arrangements and to engineer composite
particles with non-trivial internal structure. Such non close packed
supra-particles could be used as a mask for patterning or directly
as a functional component for nanoscale applications. Although we
consider only two-dimensional systems, similar behavior is expected
in a three-dimensional system of core-corona particles confined in
a spherical box.

Particles interacting through core-corona potentials represent not
only a simple model system to study numerically, such simple pair
potentials are often used to describe effective interactions among
substances with supramolecular architecture \cite{malescio3,denton}.
For example, the case of diblock copolymers, dendritic polymers, hyper-branched
star polymers, colloidal particles with block-copolymers grafted to
their surface where self-consistent field calculations lead to effective
interactions that can be modeled by a square-shoulder potential \cite{norizoe}.
Such interactions can be controlled by adjusting the length, the grafting
density of the grafted polymers, the quality of the solvent, etc.
Also, isotropic core-softened potentials have been used to qualitatively
describe the anomalous behavior of water and some other liquids and
the effects that confinement have on the phase diagram \cite{fomin}.

Numerical simulations in flat space have shown that single component
softened-core repulsive potentials may give rise to strip phases \cite{malescio1}-\cite{fornleitner1}
and periodic structures that are explained in terms of the competing
interactions between the hard core and the soft shoulder. Our aim
is to study the influence that the size of a circular confining box
has in the domain formation when the particles are restricted to move
in the interior of the confining box.

The paper is organized as follows. Section 2 presents the model and
the details for the Monte Carlo simulation. In Section 3 we show some
of the complex structures obtained, a phase diagram for particular
values of the parameters and the evolution of the assembled structures
as confinement increases. Finally, in Section 4 we present our conclusions.

\section{Methods}

\begin{figure}
\centering{}\includegraphics[viewport=50bp 0bp 800bp 618bp,scale=0.5]{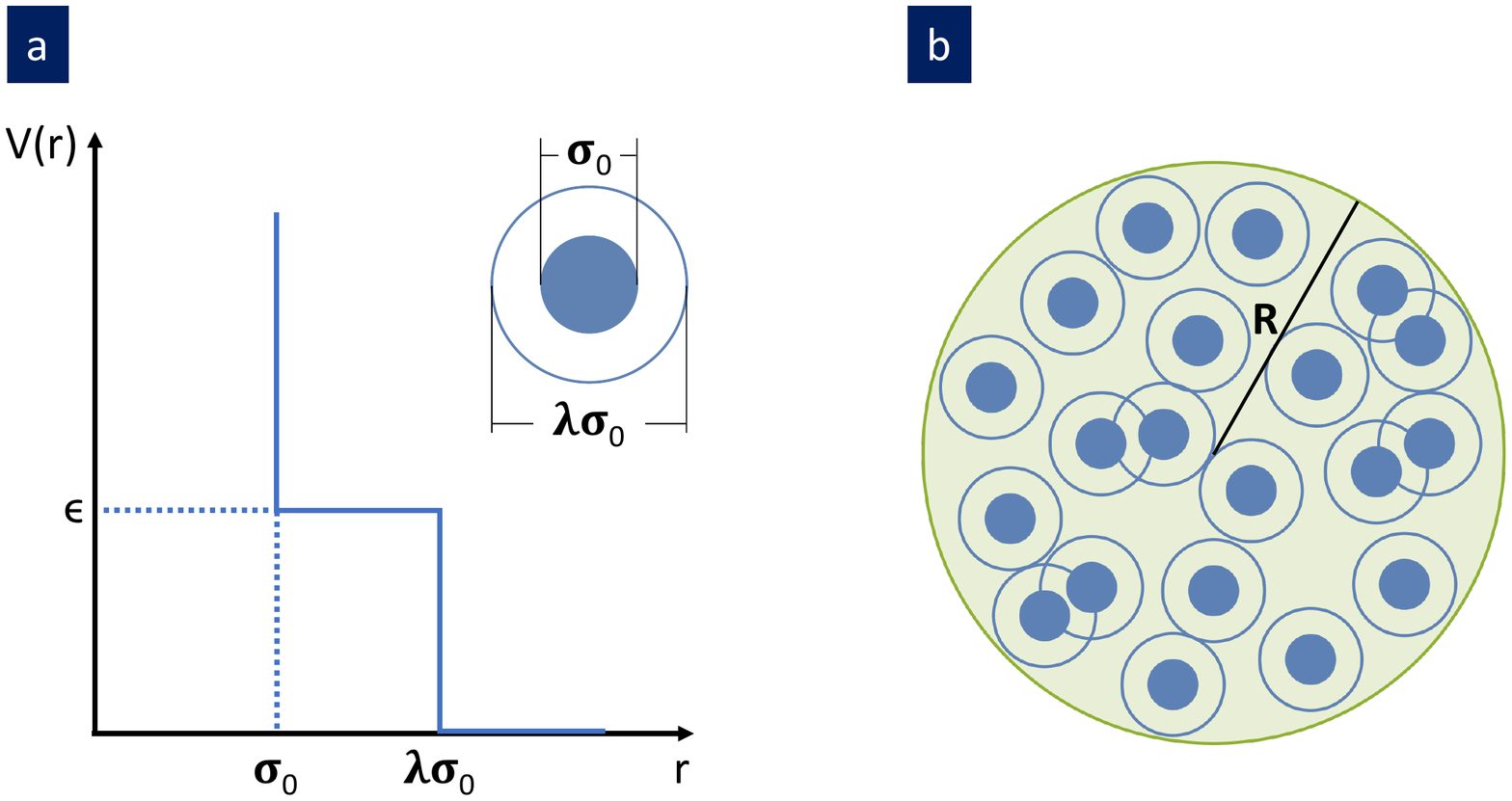}\caption{Description of the model. (a) The interaction between a pair of particles
is modeled by a hard-core soft-corona potential. (b) A collection
of particles confined in a circular box}
\label{fig model}
\end{figure}

We employ a simple model system that treats the interactions between
soft nanoparticles in an effective way via an infinite hard core encircled
by a soft repulsive shoulder. Specifically, our system consists of
particles interacting through an isotropic pair potential composed
by an impenetrable core of diameter $\sigma_{0}$ with an adjacent
square shoulder with range $\lambda\sigma_{0}$,

\begin{equation}
V\left(r\right)=\begin{cases}
\infty & r\leq\sigma_{0}\\
\epsilon & \sigma_{0}<r\leq\lambda\sigma_{0}\\
0 & \lambda\sigma_{0}<r
\end{cases},\label{potencial1}
\end{equation}
$r$ being the pair distance (see Fig. \ref{fig model}a). At large
distances, the particles do not overlap and the interaction vanishes.
The repulsive shoulder may model the steric repulsion between nanoparticles
due to the overlap of brush-like surrounding coronas. Finally, at
small separations penetration of the compact cores is very unfavorable
and gives rise to the hard-core repulsion. The simple functional form
of the interaction potential not only captures the essential features
of colloidal particles with core-corona architecture, it also offers
many computational advantages and allows to understand using simple
geometrical considerations the system's self-assembly strategy \cite{pauschenwein1,pauschenwein2}.

We consider that the nanoparticles are confined to move in a two-dimensional
(2D) plane region limited by a circular disc of radius $R$ as depicted
in Fig. \ref{fig model}b. Standard Monte Carlo simulations based
on the canonical ensemble (NVT simulations) in a circular box of radius
$R$ have been carried out using the Metropolis algorithm. We have
used $\sigma_{0}$ and $\epsilon$ as length and energy units, respectively,
set $\lambda=2$, and studied the pattern formation dependence on
$R^{*}\equiv R/\sigma_{0}$, reduced temperature $T^{\ast}\equiv k_{B}T/\epsilon$,
where $k_{B}$ is the Boltzmann's constant; and the number of particles
$N$. The structural diversity and phase behavior is explored in detail
for small number of particles $N\leq6$ although runs with other values
of $N$ are considered, with the largest value $N=400$ as an example
of a large system. In all cases, the system is first disordered at
high temperature and then brought from $T^{\ast}=1$ to a final temperature
$T^{\ast}=0.01$ through an accurate annealing procedure with steps
of $0.01$ for a given value of the box radius $R$. The radius is
progressively reduced in steps of 0.01$\sigma_{0}$ from an initial
large value, chosen such that the particles can be contained in the
box without overlapping of their coronas, up to a final value such
that no more contraction of the box is possible due to steric effects.
An auxiliary interaction potential between the particles and the confining
circle, of the form $V_{box}(r)=\epsilon/(R-\lambda\sigma_{0}/2-r)^{2}\Theta\left(0.02\sigma_{0}+r-R+\lambda\sigma_{0}/2\right)$,
with $\Theta(x)$ the Heaviside step function, is used to ensure that
the particles remain inside the box each time its radius $R$ is reduced.
An equilibration cycle consisted, for each temperature, of at least
$1\times10^{6}$ MC steps, each one representing one trial displacement
of each particle, on average. At every simulation step a particle
is picked at random and given a uniform random trial displacement
within a radius of $0.5\sigma_{0}$. The temperature vs box size phase
diagram is obtained during the annealing process as follows: for a
given box size $R$, and temperature $T$, the average energy of the
system is obtained. Then, from the average energy vs T plot, the critical
temperature where the system changes from an isotropic random configuration
to the self-assembled structure is obtained as the point where the
curve reaches a final energy value that does not change anymore under
further cooling. This is done for all $R$ starting from the largest,
where the particles do not overlap, to the more compact one.

\section{Results}

\begin{figure}
\begin{centering}
\includegraphics[viewport=50bp 0bp 800bp 418bp,scale=0.5]{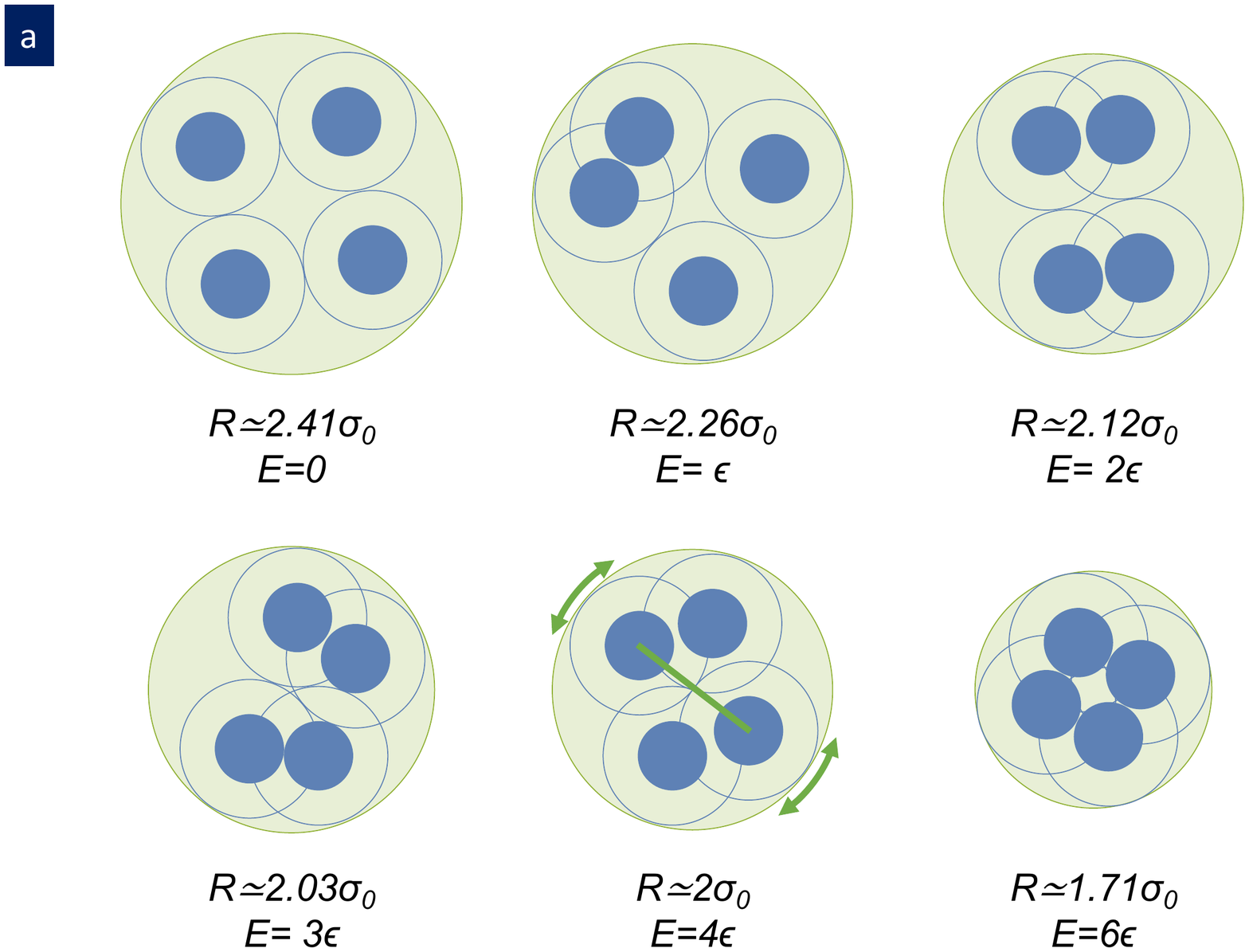}
\par\end{centering}
\centering{}\includegraphics[viewport=60bp 100bp 800bp 518bp,scale=0.5]{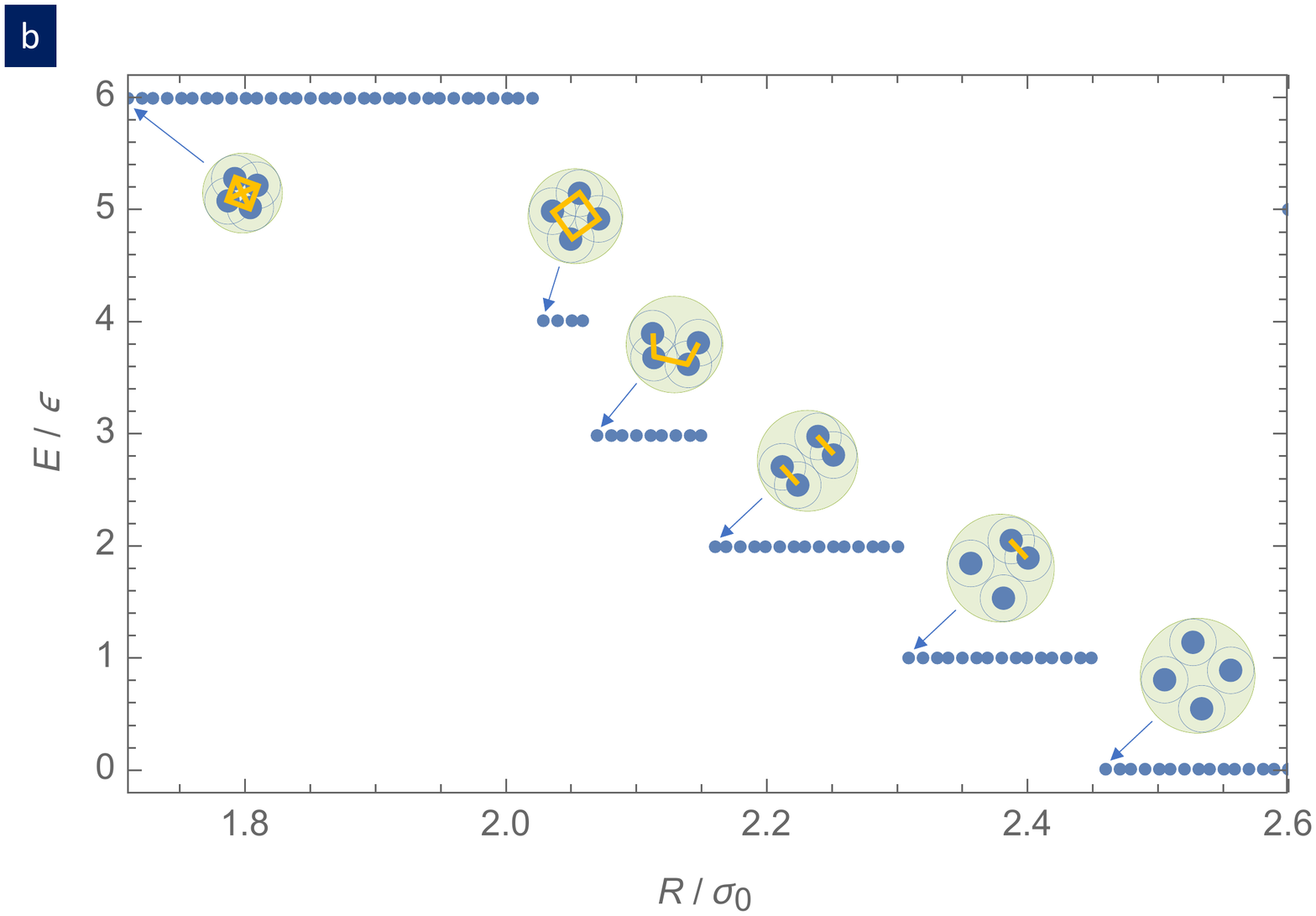}\caption{Representative minimum energy configurations for a system with four
particles with $\lambda=2$ and its energy diagram. (a) Most compact
representative configurations for given values of the interaction
energy. All the structures are rigid except the case with $E=4\epsilon$
which contain rigid dimers that move independently to each other in
a certain region. One dimer is highlighted with a green line and the
arrows indicate the movement of the dimer respect to the other. (b)
As the size of the confining box is reduced, the number of corona
overlaps increases and thus the energy increases in a stepwise fashion.
The insets show the most compact representative for a given value
of the energy. The yellow lines in the insets mark the pair of particles
that participate in a given corona overlap. Therefore, the number
of lines is the energy of the system. Notice that no configurations
with five overlarps are allowed in this example.}
\label{EvsR}
\end{figure}

In unconfined space, core-corona particles make a plethora of mesophases.
Thus, we expect to obtain a large variety of interesting structural
features for this simple model at low temperatures as it is put in
evidence in Fig. \ref{EvsR}, where a few of them are exhibited for
a confined system with a very small number of particles $N=4$, corona
to core ratio $\lambda=2$ and for different sizes of the confining
box, as indicated. Panel (a) shows the most compact representatives
of every kind of arrangement obtained in this case. Each configuration
is the result of the system minimizing its free energy and differs
from the others by the number of corona overlaps. For the larger box
size shown, the four particles adopt a configuration determined by
the particles just touching without overlapping their coronas, thus
its free energy is $E=0$. As the confining box decreases its size,
the system can no longer avoid some overlapping and will adopt the
configuration or configurations that minimize the number of corona
overlaps. The second configuration corresponds to an arrangement in
which there is only one corona overlap for which $E=\epsilon$. This
configuration with only one overlap persists under progressive shrinking
of the confining box up to a point in which this situation can no
longer be maintained and a configuration with two overlaps appears
as shown by the third case in panel (a). Further decrease of the box
size will produce progressively, configurations with three, four,
and six overlaps. The case with five overlaps is impossible due to
geometrical constraints. Notice that all the most compact configurations
shown are strictly rigid, except the case with $E=4\epsilon$, which
consists of two rigid dimers, each one consisting of a pair of particles
located at opposite sides of the box, just touching their coronas.
Each dimer can rotate independently of each other within the constrain
imposed by the steric repulsion between the cores. This is indicated
by a green line joining the particles forming one of the dimers and
the arrows indicate its relative movement with respect to the other
dimer. The presence of these ``rattlers'' near $T^{*}=0$ will depend
on the geometrical parameters of the system, $N$, $\lambda$, and
$R$. They are the result of a lack of bonds to maintain rigidity
as required by the Maxwell criteria which is a global relation that
stipulates that the number of zero modes, $n_{m}$, present in an
arbitrary mechanical structure, is simply given by the difference
between the number of degrees of freedom minus the number of independent
constraints \cite{maxwell}. For a system with $N$ particles in $d$
dimensions subjected to $N_{c}$ constraints the Maxwell criteria
reads

\begin{equation}
n_{m}-n_{ss}=dN-N_{c},\label{maxwell}
\end{equation}
where $n_{ss}$ is the number of redundant constraints or states of
self-stress. For example, the structures with $E=0,\epsilon,2\epsilon,3\epsilon,$
and $6\epsilon$ in Fig. \ref{EvsR} a) have $N_{c}=5$ constraints
and no redundant constraints, $n_{ss}=0$. Four of the constraints
originates from the direct interactions between the particles and
one from the condition that the particles are confined touching the
inner border of the common circular box. Accordingly, the Maxwell
criterion gives $n_{m}=3$ which corresponds to the three trivial
modes: two independent translations and a global rotation, thus, all
these structures are rigid. On the other hand, the structure with
$E=4\epsilon$ have $N_{c}=4$ constraints, two of them arise from
the internal bonds in the form of two identical dimers and two from
the constraint that the centers of the dimers coincide as a result
of being in a common box. Consequently, $n_{m}=4$, from which three
correspond to the trivial modes and one to a non-trivial mode corresponding
to the relative rotation of one dimer with respect to the other as
indicated by the green arrows. Although the configurations shown in
Fig. \ref{EvsR} correspond to the more compact representatives for
a given energy, other configuration with the same energy can not be
excluded. Fig. \ref{EvsR} b) shows the free energy at zero temperature
for different values of the box size. The free energy curve shows
steps that correspond to the number of corona overlaps for a given
confining box size $R$. The most compact representative configurations
for each energy are shown in the inserted drawings. The orange lines
in the insets highlights the particles that overlap their coronas.

\begin{figure}
\centering{}\includegraphics[viewport=50bp 0bp 800bp 618bp,scale=0.5]{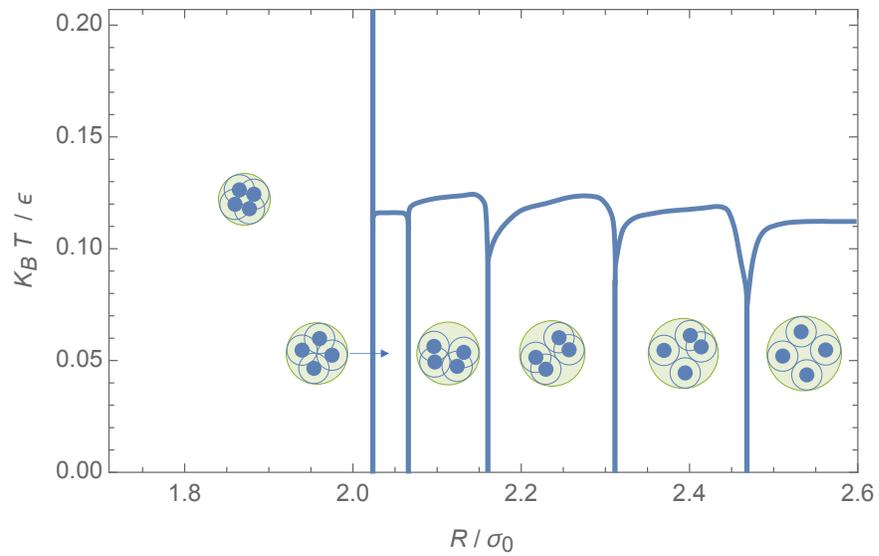}\caption{Temperature vs. box radius phase diagram for a system with four particles
and $\lambda=2$. The insets show the most compact representative
for each minimum energy configuration.}
\label{TvsR}
\end{figure}

In Fig. \ref{TvsR} we show a phase diagram in the plane temperature
vs box radius. This diagram shows that at high temperatures the particles
move inside the box randomly without adopting a particular structure,
except at small $R$. As $T$ is lowered, the system abruptly adopts
its final self-assembled structure for which the energy remains constant
under further cooling. Notice that there exists small regions showing
reentrant melting (around $k_{B}T/\epsilon\simeq0.10$) for which
an increase in $R$ may lead from an ordered structure to a fluid
like phase to another ordered structure. When the system is very small,
which for the example shown, occurs at sizes $R\lesssim2$, there
is not enough room for the particles to move more or less freely since
they are sterically constrained by the presence of the hard cores
of the other particles, and the confining box. This is a reminiscence
of the athermal behavior of hard discs since no change in the number
of corona overlaps is possible at any temperature for these box sizes.

\begin{figure}
\centering{}\includegraphics[viewport=150bp 50bp 800bp 618bp,angle=-90,scale=0.7]{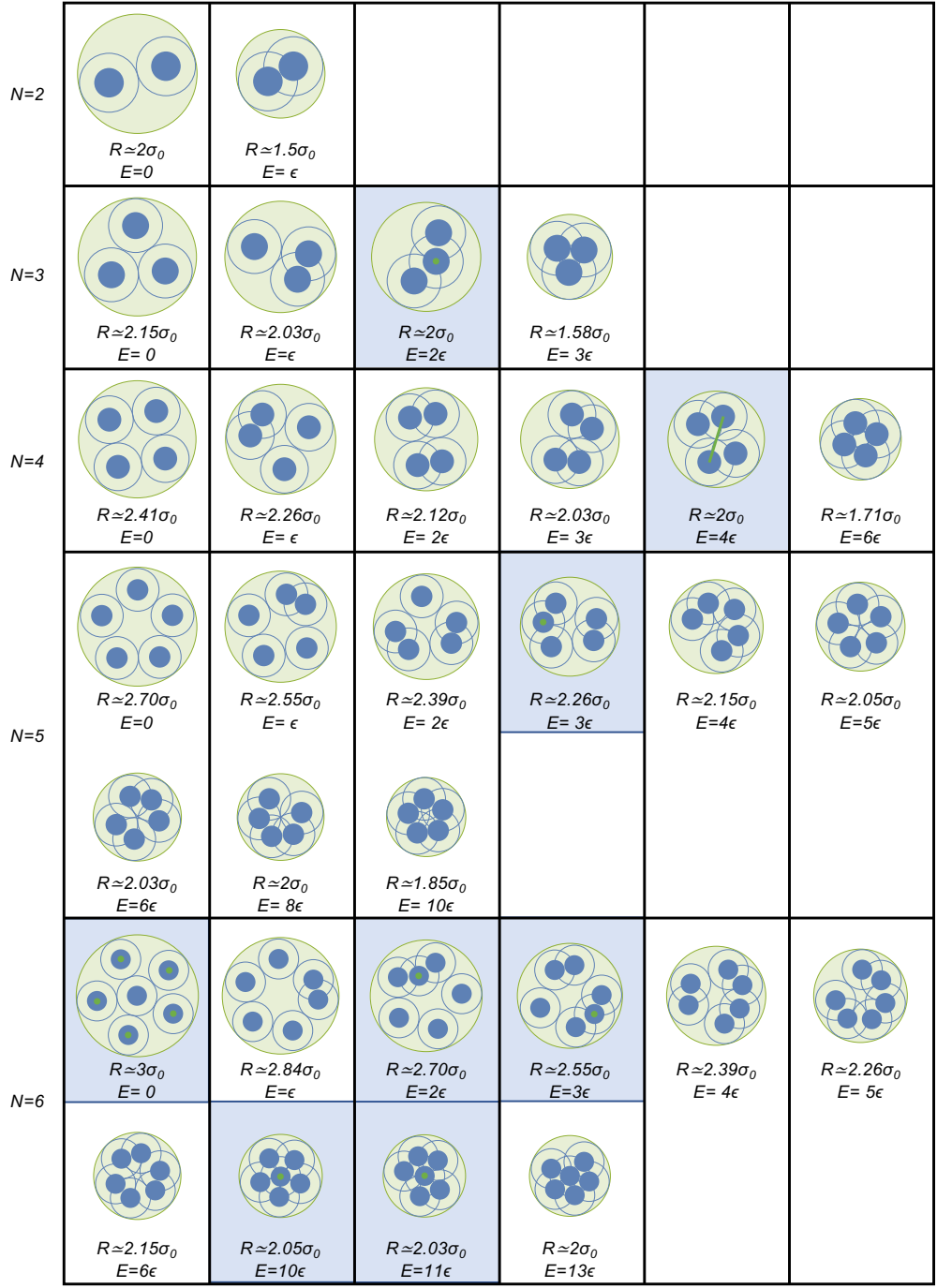}\caption{Representative minimum energy configurations for systems with a small
number of particles and $\lambda=2$. The number of configurations
grows as the number of particle increases. Blue panels indicate non-rigid
structures and the particles highlighted with a green dot are rattlers.}
\label{Configurations}
\end{figure}

In Fig. \ref{Configurations}, we identify the most compact representatives
of the minimum energy configurations for systems with $\lambda=2$
and $N\leq6$. The energy of each configuration is indicated as well
as the theoretical values of the box radius corresponding to each
compact representative, as obtained by geometrical considerations.
As expected, the structural complexity increases with $N$. Some of
these configurations correspond to rigid structures, while others
present rattlers, that is, particles that may change their relative
positions to other particles in spite of being the more compact representative
for its energy, while maintaining the energy of the configuration
constant. To mention just a few of the non-rigid structures, for $N=3$,
the third configuration shows a rattler while for $N=6$, the first,
third, fourth, eight, and nineth configurations show rattlers. Rattlers
are highlighted by green dots (or a green line in the case of a dimer)
and non-rigid structures are indicated by the blue panels. Notice
that the configurations with $E=0$ are analogous to those corresponding
to confined hard disks with diameter $\lambda\sigma_{0}$. One may
expect the same equivalence would be observed for the most compact
cases as compared to hard disks with diameter $\sigma_{0}$. This
is true for the cases with $N=1$ to $5$ but for $N=6$ the most
compact configuration has no equivalent in the hard-disks case. This
is due to the fact that the coronas of particles located on opposite
sides of the confining box provide extra constraints that avoid the
peripheral disks to rotate in contrast with the case with $N=6$ and
$E=0$.

\begin{figure}
\begin{centering}
\includegraphics[viewport=0bp 0bp 800bp 418bp,scale=0.5]{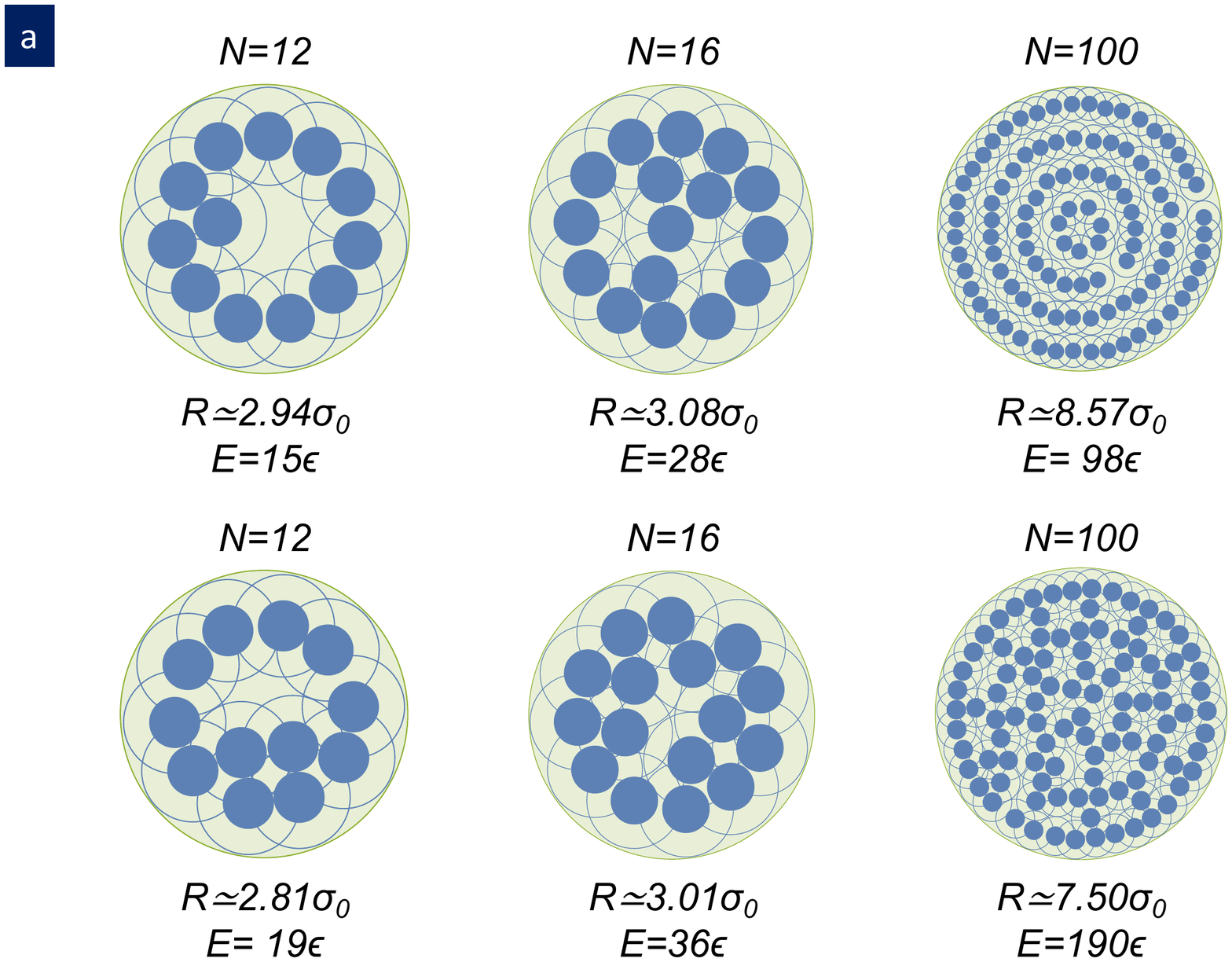}
\par\end{centering}
\centering{}\includegraphics[viewport=0bp 0bp 800bp 518bp,scale=0.5]{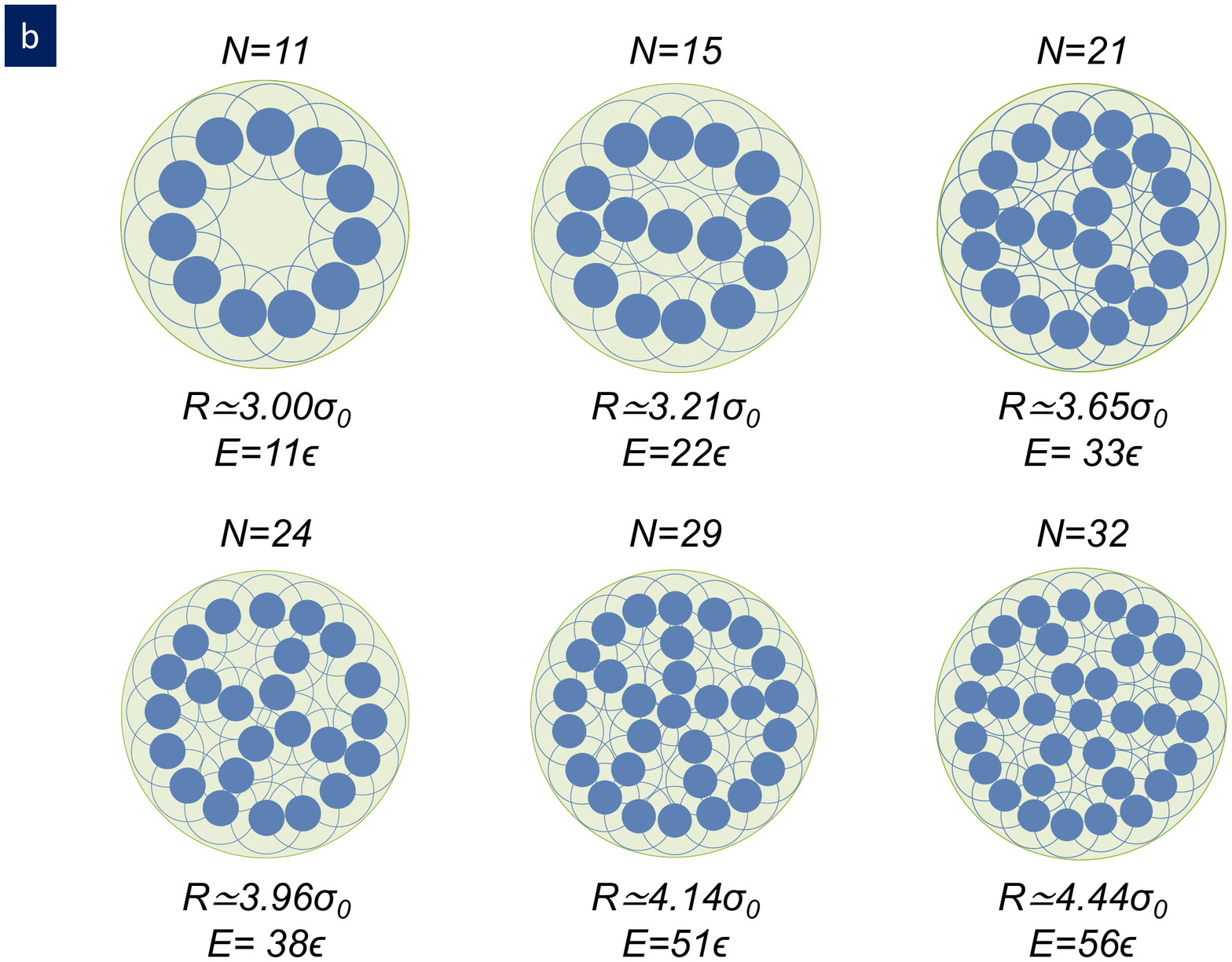}\caption{Other interesting configurations for systems with $\lambda=2$. (a)
Two representative configurations for each of $N=12$, $N=16$, and
$N=100$ particles. (b) Structured \textit{microparticles} with a
progressive number of \textit{compartments}.}
\label{Other}
\end{figure}

Low energy configurations for other values of $N$ larger than $6$
have also been considered, however, given the large variety and the
complexity of the structures arising upon increasing $N$, we have
not attempted to enumerate them. In fact, more sophisticated algorithms
than the one used in this work would be necessary to thoroughly explore
the energy landscape in search of a global minimum. In Fig. \ref{Other}
we show interesting representatives for larger values of $N$, as
indicated. For large number of particles we observe the formation
of concentric shells of overlapping disks. This behavior is similar
to the formation of concentric shells obtained in the case of softly
repulsive particles confined in a spherical box \cite{mughal}. As
the radius of the confining box decreases, this shells are intercalated
with approximately uniformly distributed particles as seen in the
last figure of the top panel. Residual defects are seen in this figure.
It is possible to control to certain extent the resulting structure
by an appropriate choice of the number of particles, the width of
the corona, and the radius of the confining box as shown in the lower
panel of Fig. \ref{Other}. Here, the parameters where chosen specifically
in order to obtain a sequence of configurations with different rotational
symmetry or more precisely, with an increasing number of \textit{compartments},
from one to six.

\begin{figure}
\begin{centering}
\includegraphics[viewport=50bp 0bp 800bp 618bp,scale=0.5]{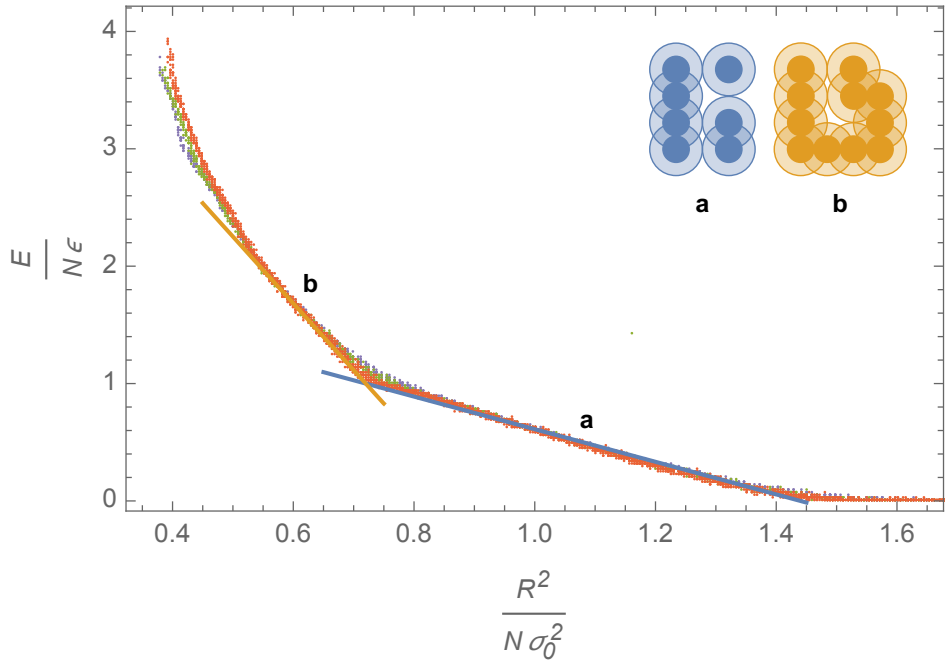}
\par\end{centering}
\centering{}\caption{Pathway towards close packing. As the size of the confining box is
reduced, the number of corona overlaps increases and thus the energy
increases in a piecewise fashion as a consequence of the consecutive
sequence of progressively denser and more branched configurations.
An schematics of the first two sequential configurations present are
shown in the inset. Symbols are simulation result with $N=100$ (orange),
$N=200$ (green), and $N=400$ (purple). The lines are the results
of the models given by Eqs. (\ref{E}) (blue) and (\ref{E2}) (yellow).}
\label{EvsR-100}
\end{figure}

Figure \ref{EvsR-100} shows the free energy vs box size for large
systems ($N=100,200,$ and $400$) in the normalized axis, $E/(N\epsilon)$
vs $R^{2}/(N\sigma_{0}^{2})$. The symbols correspond to simulation
results. Note that in these axis all the curves collapses in a single
universal curve and this curve presents piecewise behavior with approximately
linear sections. Different regions correspond to different arrangements
of the particles. For example, for large $R$, the box is so large
that no overlaps between coronas exist and the energy of the system
is zero. Then, at $R^{2}/(N\sigma_{0}^{2})\simeq1.44$ the number
of corona overlaps increase progressively as $R$ decreases. The inset
labeled (a) shows schematically a group of particles, some of them
overlap with the nearest neighbors and some other do not overlap,
forming lines similar to the ones shown for the case $N=100$ and
$E=98\epsilon$ in Fig. \ref{Other} (a). A simple model can be proposed
to describe the behavior in this regime. Let us imagine that the particles
form a straight line of length $L=N_{0}\lambda\sigma_{0}+N_{1}\sigma_{0}$
and width $W=\lambda\sigma_{0}$, where $N_{0}$ the number of particles
that do not overlap and $N_{1}$ the number of particles that overlap
with their neighbors, so that $N=N_{0}+N_{1}$. The area fraction
$\phi_{1}\left(N_{1}\right)$ of disks with diameter $\lambda\sigma_{0}$
in the rectangle of area $WL$ is

\begin{equation}
\phi_{1}\left(N_{1}\right)=\frac{N\frac{\pi\lambda^{2}\sigma_{0}^{2}}{4}}{\alpha WL}=\frac{N\pi\lambda}{4\alpha\left[\lambda N+\left(1-\lambda\right)N_{1}\right]},\label{phi}
\end{equation}
where $\alpha$ is a factor that takes into account the fact that
the disks can not cover completely the rectangular stripe. This region
starts at $\phi_{1}\left(N_{1}=0\right)\equiv\phi_{0}$, with $\phi_{0}$
the maximum packing of disks without overlaps. This quantity is then

\begin{equation}
\phi_{0}=\frac{N\frac{\pi\lambda^{2}\sigma_{0}^{2}}{4}}{\pi R_{0}^{2}},\label{phi0}
\end{equation}
where $R_{0}$ is the radius of the confining box when the overlaps
start to appear. Using Eqs. (\ref{phi}) and (\ref{phi0}) we get

\begin{equation}
\phi_{1}\left(N_{1}\right)=\frac{N^{2}\lambda^{3}\sigma_{0}^{2}}{4R_{0}^{2}\left[\lambda N+\left(1-\lambda\right)N_{1}\right]}.\label{phi1}
\end{equation}
In general

\begin{equation}
\phi_{1}\left(N_{1}\right)=\frac{N\frac{\pi\lambda^{2}\sigma_{0}^{2}}{4}}{\pi R^{2}},\label{phi2}
\end{equation}
so that using Eqs. (\ref{phi1}) and (\ref{phi2}) and since $E=N_{1}\epsilon$
then

\begin{equation}
\frac{E}{N\epsilon}=\frac{\lambda}{\lambda-1}\left(1-\frac{R^{2}}{R_{0}^{2}}\right),\label{E}
\end{equation}
where $\lambda\leq2$ is being assumed. This regime ends at $R_{1}$,
when $N_{1}=N$, so that all particles overlap and further shrinking
of the box will produce more compact configuration like lines with
kinks or branching points. From Eq. (\ref{E}) we get $R_{1}=R_{0}/\sqrt{2}$.
From the simulation results, $R_{0}^{2}/\left(N\sigma_{0}^{2}\right)\simeq1.44$,
that once inserted in the model given by Eq. (\ref{E}) and using
$\lambda=2$ gives the blue line in Fig. \ref{EvsR-100}. Thus, the
model reproduces correctly the simulation results.

The next regime consists of $N_{1}$ particles that overlap with their
next neighbors in the line and $N_{2}$ kinks or branching points
that has an extra overlap as shown in inset (b) of Fig. \ref{EvsR-100}.
The total number of particles is $N=N_{1}+N_{2}$. Proceeding as before,
let us assume that the particles are contained in a rectangular region
of length $L=N_{1}\sigma_{0}+\mu\sigma_{0}N_{2}$ and width $W=\lambda\sigma_{0}$,
with $\mu$ a fitting parameter that has to be introduced since kinks
contribute less to the length of the rectangle such that $\mu<1$.
As before, the area fraction $\phi_{2}\left(N_{2}\right)$ of disks
in the rectangle of area $WL$ is

\begin{equation}
\phi_{2}\left(N_{2}\right)=\frac{N\frac{\pi\lambda^{2}\sigma_{0}^{2}}{4}}{\alpha_{2}WL}=\frac{N\pi\lambda}{4\alpha_{2}\left[N+\left(\mu-1\right)N_{2}\right]},\label{phi3}
\end{equation}
where again $\alpha_{2}$ takes into account the fact that the rectangular
stripe can not be completely covered by the disks. From the condition
$\phi_{2}\left(N_{2}=0\right)=\phi_{1}\left(N\right)$ we get $\alpha_{2}=\alpha$
and

\begin{equation}
\phi_{2}\left(N_{2}\right)=\frac{N^{2}\lambda\sigma_{0}^{2}}{\left[N+\left(\mu-1\right)N_{2}\right]R_{0}^{2}}.\label{phi4}
\end{equation}
Now, the free energy $E=\left(N_{1}+2N_{2}\right)\epsilon=\left(N+N_{2}\right)\epsilon$,
and using Eq. (\ref{phi2}), with $\phi_{2}\left(N_{2}\right)$ replacing
$\phi_{1}\left(N_{1}\right)$ and Eq. (\ref{phi4}) we get

\begin{equation}
\frac{E}{N\epsilon}=1+\frac{1}{1-\mu}\left(1-\frac{4R^{2}}{\lambda R_{0}^{2}}\right),\label{E2}
\end{equation}
where again, $\lambda\leq2$ is being assumed. In Fig. \ref{EvsR-100}
we show with yellow line (b) the predictions of Eq. (\ref{E2}) with
$R_{0}^{2}/\left(N\sigma_{0}^{2}\right)\simeq1.44$, $\lambda=2$,
and the fitting parameter set to $\mu\simeq0.76$. The rest of the
curve grows more rapidly due to the increase in the number and the
coordination number of the branching points as the box shrinks further.

\section{Conclusions}

In this paper we explored the phase behavior of model colloidal particles
interacting via a core-corona potential confined inside a circular
box. Specifically, we analyze how structural diversity depends on
the size of the confining box for a particular choice of the potential
parameters. For small number of particles ($N\leq6$), we identify
the configurations establishing global energy minima and compare the
results to the global minima corresponding to hard disks. We show
that unlike the hard disks, for the same number of particles the system
presents more than one global minimum depending on the radius of the
confining cage and that in some cases the minimum-energy configurations
contain rattlers, that is, particles that can move freely in a restricted
region without changing the global minimum energy. Also, we have found
that the presence of the coronas, confers stability to configurations
that are not possible for hard disks. Interestingly, we obtain a large
variety of open structures with diverse degree of anisotropy in spite
of the isotropy of the interactions and of the confining box.

For large number of particles, the energy vs box radius curve shows
the presence of well defined regimes characterized by the progressive
appearing of branching points. Large confining boxes may allocate
the particles without shoulder overlaps. As the size of the box decreases
small strings start to appear up to the point of forming concentric
rings without branches. At a certain point the box is so small that
the particles can not be allocated unless branching points appear.
The number of the branching points and the number of corona overlaps
increase as the box size decreases finally reaching the most compact
possible packing where all the particles form part of branching points.

The resulting assemblies can be considered as micro-structured particles
with varying degree of complexity. Such particles suitably ordered
on a surface can be used as a mask for patterning or directly as functional
components for more complex structures.

{\LARGE{}Acknowledgements}{\LARGE\par}

We acknowledge partial financial support provided by DGAPA-UNAM through
grant DGAPA IN-103419.

\newpage{}

\end{document}